\documentclass[aps,a4paper, preprint, superscriptaddress,preprintnumbers,floatfix,
nofootinbib,amsmath,amssymb]{revtex4}
\usepackage{graphicx}
\usepackage{epsfig}
\usepackage{amsmath}
\usepackage{amsfonts}
\usepackage{amssymb}
\usepackage{url}
\usepackage{hyperref}
\usepackage{subfigure}
\usepackage{hhline}
\usepackage{bm}
\usepackage{cancel}
\usepackage[toc,page]{appendix}
\usepackage{comment}
\usepackage{hyperref}
\usepackage{amsmath}
\usepackage{graphicx}
\usepackage[usenames]{color}
\usepackage{minitoc}
\usepackage{tabulary}
\usepackage{array}
\usepackage{setspace}
\usepackage{enumerate}
\usepackage{acronym}
\usepackage{multirow}
\usepackage{url}
\usepackage{xcolor,listings} 
\usepackage{natbib}
\begin{document}
\title{Deviation from Tribimaximal mixing using $A_{4}$ flavour model with five extra scalars}

\author{Victoria Puyam}
\email{victoria.phd.phy@manipuruniv.ac.in}
 \affiliation{Department of Physics, Manipur University, Imphal-795003, India}                                      
  \author{ S. Robertson Singh}
\email{robsoram@gmail.com}
 \affiliation{Department of Physics, Manipur University, Imphal-795003, India}
  \author{ N. Nimai Singh}
\email{nimai03@yahoo.com}
 \affiliation{Department of Physics, Manipur University, Imphal-795003, India}   
   \affiliation{  Research Institute of Science and Technology, Imphal-795003, India }   

%%%%%%%%%%%%%%%%%%%%%%%%%%%%%%%%%%%%%%%%%%%%%%%%%%%
\begin{abstract}
%%%%%%%%%%%%%%%%%%%%%%%%%%%%%%%%%%%%%%%%%%%%%%%%%%%%
A  modified neutrino mass model with five extra scalars  is constructed using $A_{4}$ discrete symmetry group.  The resultant mass matrix  is able to give necessary deviation from Tribimaximal mixing  which reproduces the current neutrino masses and mixing data with good accuracy. The model gives testable prediction  for the future  measurements of the neutrinoless double-beta decay parameter $|m_{\beta \beta}|$. The analysis is consistent with latest cosmological bound $\Sigma m_{i}\leq$ 0.12 eV.
\\


\end{abstract}

\maketitle

\section{Introduction}

In the last few decades, the neutrino experiments have confirmed neutrino oscillations and mixings through the observation of solar and atmospheric neutrinos, indicating their masses thereby providing an important solid clue for a new physics beyond the Standard Model (SM) of particle physics. At present, the neutrino oscillation experiments \cite{KamLAND:2002uet, SNO:2002tuh, Super-Kamiokande:1998kpq, DoubleChooz:2011ymz, Lasserre:2012ax, Ling:2013fta, McDonald:2016ixn} have measured the oscillation parameters viz: mass squared differences ($\Delta m^{2}_{21}$ and $\Delta m^{2}_{31})$ and  mixing angles $(\theta_{12}, \theta_{23}$ and $\theta_{13})$ to a good  accuracy. The bounds on the absolute neutrino masses scale are also greatly reduced by direct neutrino mass experiments \cite{KATRIN:2019yun}, the neutrinoless double beta decay experiments ($0\nu \beta \beta)$ \cite{GERDA:2019ivs, CUORE:2019yfd, KamLAND-Zen:2016pfg} and cosmological observation \cite{ Planck:2018nkj}. However, the current data is still unable to explain several key issues such as the octant of $ \theta_{23}$, the neutrino mass ordering, CP violating phase, etc.

  The oscillation data reveals  certain pattern of neutrino mixing matrix. Out of the  several approaches to explain the observed pattern, the Tribimaximal mixing (TBM) \cite{Harrison:2002er, Xing:2002sw} used to be very favourable.  The $A_4$ flavour symmetry model proposed by Altarelli and Feruglio \cite{Alta2005yp,Altarelli:2010gt} can accommodate TBM mixing scheme in a neutrino mass model. The TBM mixing pattern has the form
  \begin{align*}
  U_{TBM}=\left( \begin{array}{ccc}
			\sqrt{\frac{2}{3}}&\sqrt{\frac{1}{3}} & 0\\
  -\sqrt{\frac{1}{6}}&\sqrt{\frac{1}{3}}& -\sqrt{\frac{1}{2}}\\
   -\sqrt{\frac{1}{6}}&\sqrt{\frac{1}{3}} & \sqrt{\frac{1}{2}}
			\end{array}\right).
\end{align*}
 However, the recently observed  non-zero  $\theta_{13}$  disfavours TBM and leads to the modification of several mass models constructed with TBM \cite{Bjorken:2005rm, He:2006qd, Gupta:2011ct, Kumar:2010qz, Borah:2017dmk, Das:2018qyt}.  As a result, the neutrino mixing patterns like TM1 \cite{Rodejohann:2012cf} and TM2 \cite{Albright:2008rp,Li:2022bqy} which are proposed with slight deviation from TBM, gain momentum. Currently, they can predict the observed pattern and mixing angles with good consistency. 
  
 In this present work, we propose a model with five extra SM singlet scalars to explain  neutrino  parameters in their experimental ranges. The present model is constructed in the basis where charged lepton is diagonal. The deviation from TBM and  non-zero value of $\theta_{13}$ are obtained as a consequence of specific Dirac mass matrix which is constructed using antisymmetric part arising from the product of two $A_{4}$ triplets. Here, we present a detailed analysis on the neutrino oscillation parameters and its correlation among themselves and with  neutrinoless double-beta decay parameter $|m_{\beta \beta}|$. 
 
 The paper is organized as follows: In Section 2, we present the description of the model with the particle contents in the underlying symmetry group.  In Section 3, we give the detailed numerical analysis and the results in terms of correlation plots.  Section 4  deals with   summary and conclusion. The Appendix   provides  brief description of $A_{4}$ discrete group.  
   
\section{ Description of the model}

 We extend  the SM by adding five extra scalars namely $\xi_{1}$, $\xi_{2}$, $\xi_{3}$, $\phi_{T}$ and $\phi_{S}$ which are transformed as  1, $1^{''}$, $ 1^{'}$, 3 and 3 respectively under $A_{4}$ group. The SM lepton doublet l are assigned to the triplet representation under $A_{4}$, right-handed charged lepton $e^c$, $\mu^c$, $\tau^c$, and right-handed neutrino field $\nu^{c}$ are assigned to the $A_{4}$ representations 1, $1^{''}$, $ 1^{'}$ and 3, respectively. The right-handed neutrino field $\nu^{c}$ contributes to the effective neutrino mass matrix through Type-I see-saw mechanism. The $A_{4}$ symmetry is supplemented by additional $Z_{3}\times Z_{2}$ group to restrict additional terms otherwise allowed by $A_{4}$ symmetry. The transformation properties of the fields used in the model are given in  Table \ref{t1}.
   
\begin{table}[!h]
\centering
\begin{tabular}{|m{2cm}|m{1cm}|m{1cm}|m{1cm}|m{1cm}|m{1cm}|m{1cm}|m{1cm}|m{1cm}|m{1cm}|m{1cm}|m{1cm}|m{1cm}|m{1cm}|}
\hline 
~~Fields &~~ $l$ &~~ $e^{c}$ &~~$\mu^{c}$  &~~$\tau^{c}$&~~$\nu^{c}$  &~~$H_{u,d}$ &~~$\phi_{S}$&~~$\phi_{T}$ &~~$\xi_{1}$&~~$\xi_{2}$&~~$\xi_{3}$ \\ 
\hline 
~~$A_{4}$ &~~3& ~~1&~~$1^{''}$ & ~~$1^{'}$&~~3&~~1 &~~3 &~~3& ~~1&~~$1^{''}$&~~$1^{'}$ \\ 
\hline 
~~$Z_{3}$ &~~$\omega^{2}$ &~~$\omega$ &~~$\omega$ &~~$\omega$&~~1 &~~1 &~~$\omega$ &~~1 &~~$\omega$ &~~$\omega$&~~$\omega$\\
\hline
~~$Z_{2}$ &~~1 &~-1 &~-1 &~-1&~~1 &~~1 &~~1 &~-1 &~~1 &~~1&~~1\\
\hline
~~$SU(2)_{L}$& ~~2&~~1&~~1&~~1&~~1&~~2&~~1&~~1&~~1&~~1~~&~~1\\
\hline
\end{tabular}
\caption{Transformation properties of various fields under $A_{4}\times Z_{3} \times Z_{2}\times SU(2)_{L}$ group.}
\label{t1}
\end{table}

The Yukawa Langrangian terms for the leptons which are invariant under $A_{4}\times Z_{3} \times Z_{2}\times SU(2)_{L}$ transformation, are given in the equation:
\begin{align}
\label{e1}
-L_{Y}=& \frac{Y_{e}}{\Lambda}(l\phi_{T})_{1}H_{d}e^{c}+\frac{Y_{\mu}}{\Lambda}(l\phi_{T})_{1^{'}}H_{d}\mu^{c}+\frac{Y_{\tau}}{\Lambda}(l\phi_{T})_{1^{''}}H_{d}\tau^{c} \nonumber \\
& + \frac{ y_{1}}{\Lambda}\xi_{1}(lH_{u}\nu^{c})_{1}+\frac{y_{2}}{\Lambda}(\xi_{2})_{1''}(lH_{u}\nu^{c})_{1'}+ \frac{y_{3}}{\Lambda}(\xi_{3})_{1'}(lH_{u}\nu^{c})_{1''} \nonumber \\
& + \frac{y_{a}}{\Lambda}\phi_{S}(lH_{u}\nu^{c})_{A}+\frac{y_{b}}{\Lambda}\phi_{S}(lH_{u}\nu^{c})_{S}+\frac{1}{2}M_{N}(\nu^{c}\nu^{c})+h.c.,
\end{align}

\begin{table}[!h]
\begin{center}
\begin{tabular}{ccc}
\hline 
  Fields &  &Vacuum Expectation Value(VEV)  \\
  \hline 
$<\phi_{S}>$  & &$(v_{s}, v_{s},v_{s})$\\ 
 
$<\phi_{T}>$ &  & $(v_{T},0,0)$ \\ 

$<H_{u}>$,$<H_{d}>$ &  &$v_{u}$, $v_{d}$  \\ 
  
$<\xi_{1}>$, $<\xi_{2}>$, $<\xi_{3}>$ &  & $u_{1}$, $u_{2}$, $u_{3}$ \\ 
\hline
\end{tabular}
\end{center}
\caption{Vacuum Expectation Values (VEV) of the scalar fields used in the model.}
\label{t2} 
\end{table}

 where $\Lambda$ is the model cutoff high scale. The Yukawa mass matrices can be derived from Eq.~(\ref{e1}) by using the vacuum expectation value given in Table \ref{t2}  . The charged lepton mass matrix thus obtained, is diagonal and has the form
\begin{equation}
M_{l}=\frac{v_{d}v_{T}}{\Lambda}\begin{pmatrix}
 Y_{e}&0&0\\
 0&Y_{\mu}&0\\
 0&0&Y_{\tau}
 \end{pmatrix}.
\end{equation}

The Majorana neutrino mass matrix has the structure
\begin{equation}
M_{R}=\begin{pmatrix}
M_{N}& 0& 0\\0&0 & M_{N}\\ 0& M_{N}& 0
\end{pmatrix}.
\end{equation}

 The  Dirac mass matrix is in the form
\begin{equation}
M_{D}=\begin{pmatrix}
2 a + c& -a + b + d& -a - b + e\\ -a - b+d  & 
 2 a + e& -a + b + c\\ -a + b + e & -a - b + c& 2 a + d
\end{pmatrix}.
\end{equation}
where $a= y_{b}.v_{u}.v_{s}/ \Lambda$,  $b= y_{a}.v_{u}.v_{s}/ \Lambda$ and c, d, e = $y_{i}.v_{u}.u_{i}/ \Lambda$, \ $i=1, 2$ and 3.

The effective neutrino mass matrix is obtained by using Type-I see-saw mechanism,
\begin{align}
m_{\nu}=&M_{D}^{T}M_{R}^{-1}M_{D}
\label{e4} \\
=&\begin{pmatrix}
m_{11}&m_{12}&m_{13}\\
m_{12}&m_{22}&m_{23}\\
m_{13}&m_{23}&m_{33}
\end{pmatrix}, 
\end{align}
where
\begin{align*}
m_{11} &=\frac{1}{M_{N}}((2 a + c)^2 + 2 (a + b - d)(a - b - e))\\
 m_{12}&=m_{21}=\frac{1}{M_{N}}(-3 a^2 + b^2 +  2 c d + e^2 + b (-d + e) + a (6 b - 2 c + d + e))\\
  m_{13}&=m_{31}=\frac{1}{M_{N}} (-3 a^2 + b^2 + d^2 + 2 c e + b (-d + e) + a (-6 b - 2 c + d + e))\\ 
  m_{22} &= \frac{1}{M_{N}}((-a + b + d)^2 - 2 (a + b - c) (2 a + e))\\
  m_{23}&=m_{32}= \frac{1}{M_{N}}(6 a^2 - 2 b^2 + c^2 + 2 d e + b (-d + e) +  a (-2 c + d + e))\\ 
  m_{33}&= \frac{1}{M_{N}}2 ((-a + b + c) (2 a + d) + (a + b - e)^2)).
\end{align*}

In order to explain smallness of active neutrino masses, we consider the heavy neutrino with  masses ($M_{N})\sim \mathcal{O}(10^{15}$ GeV).  When $\text{d}=\text{e}$ and $\text{b}=\text{c}=0$, the mass matrix $m_{\nu}$ obtained in Eq.~(\ref{e4}), takes the form

\begin{equation}
\label{q1}
m_{\nu}=\frac{1}{M_{N}}\begin{pmatrix}
4 a^2 + 2 (a - e)^2& -3 a^2 + 2 a e + e^2&-3 a^2 + 2 a e + e^2)\\ 
-3 a^2 + 2 a e + e^2 &(-a + e)^2 - 2 a (2 a + e)& 6 a^2 + 2 a e + 2 e^2\\ 
-3 a^2 + 2 a e + e^2& 6 a^2 + 2 a e + 2 e^2
  & (a - e)^2 - 2 a (2 a + e)
\end{pmatrix},
\end{equation}
 with mass eigenvalues $m_{1}=(3 a - e)^2/M_{N}$, $m_{2}=(4 e^2)/M_{N}$ and  $m_{3}=-(3 a + e)^2/M_{N}$. However, in this case, the mass matrix  $m_{\nu}$ can be diagonalized by $U_{TBM}$ and this possibility would violate the currently observed neutrino oscillation data especially  the non-zero values of $\theta_{13}$. Therefore, we need to consider the non-zero values for b, c, d and e to get  $\theta_{13}\neq 0$. Hence, the  unitary matrix that can diagonalize  the resultant mass matrix, should deviate from the $U_{TBM}$. 
\section{Numerical Analysis and Results}

  \begin{table}[!h]
  \begin{center}
\begin{tabular}{cccc}
\hline
\hline 
 Parameters& Best fit$\pm $1$\sigma$& 2$\sigma$ &3$\sigma$ \\
\hline
\hline
\vspace*{0.2 cm}
$\theta_{12}/^{\circ}$ &$ 34.3\pm 1.0$& 32.3-36.4&31.4-37.4   \\  
\vspace*{0.2 cm}
$\theta_{13}/^{\circ}$(NO) &$8.53^{+0.13}_{-0.12}$&8.27-8.79& 8.20-8.97   \\ 
\vspace*{0.2 cm}
$\theta_{13}/^{\circ}$(IO) &$8.58^{+0.12}_{-0.14}$&8.30-8.83 &8.17-8.96   \\ 
\vspace*{0.2 cm}
$\theta_{23}/^{\circ}$(NO) &$49.26\pm 0.79$&47.35-50.67& 41.20-51.33   \\ 
\vspace*{0.2 cm}
$\theta_{23}/^{\circ}$(IO) &$49.46^{+ 0.60}_{-0.97}$&47.35-50.67& 41.16-51.25   \\ 
\vspace*{0.2 cm}
 $\Delta m_{21}^{2}[10^{-5}eV^{2}]$& $7.50^{+ 0.22}_{-0.20} $ &7.12-7.93& 6.94-8.14\\ 
\vspace*{0.2 cm} 
$|\Delta m_{31}^{2}|[10^{-3}eV^{2}]$(NO) &$2.55^{+0.02}_{-0.03}$& 2.49-2.60&2.47-2.63 \\
\vspace*{0.2 cm} 
 $|\Delta m_{31}^{2}|[10^{-3}eV^{2}]$(IO) &$2.45^{+0.02}_{-0.03}$&2.39-2.50&2.37-2.53 \\
\vspace{0.2 cm}
$\delta/^{\circ}$(NO)& $194^{\footnotesize{+24}}_{\footnotesize{-22}}$&152-255 & 128-359\\
\vspace*{0.2 cm}
$\delta/^{\circ}$(IO)& $284^{\footnotesize{+26}}_{\footnotesize{-28}}$ & 226-332&200-353\\
\hline
\hline
\end{tabular}
\caption{The global-fit result for neutrino oscillation parameters \cite{deSalas:2020pgw}.}
\label{t3}
 \end{center}
\end{table} 

  \begin{figure}[!h]
\centering
\subfigure[]{
\includegraphics[width=.40\textwidth]{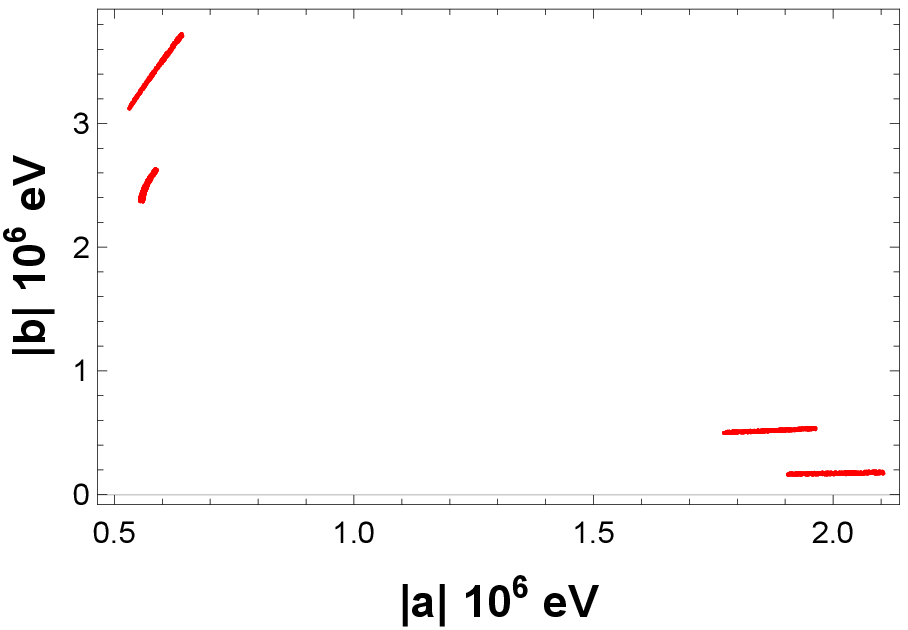}}
\quad
\subfigure[]{
\includegraphics[width=.40\textwidth]{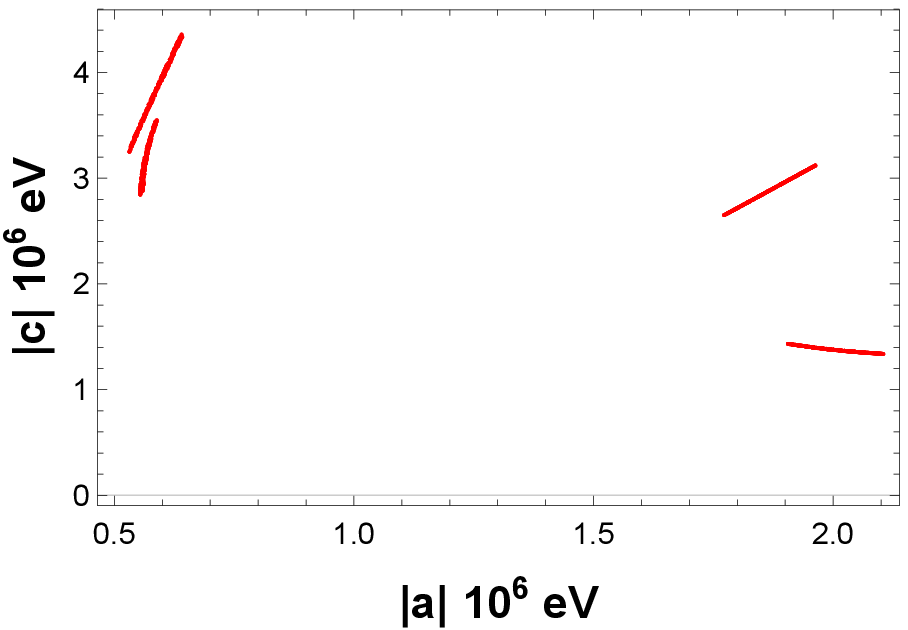}}
\quad
\subfigure[]{
\includegraphics[width=.40\textwidth]{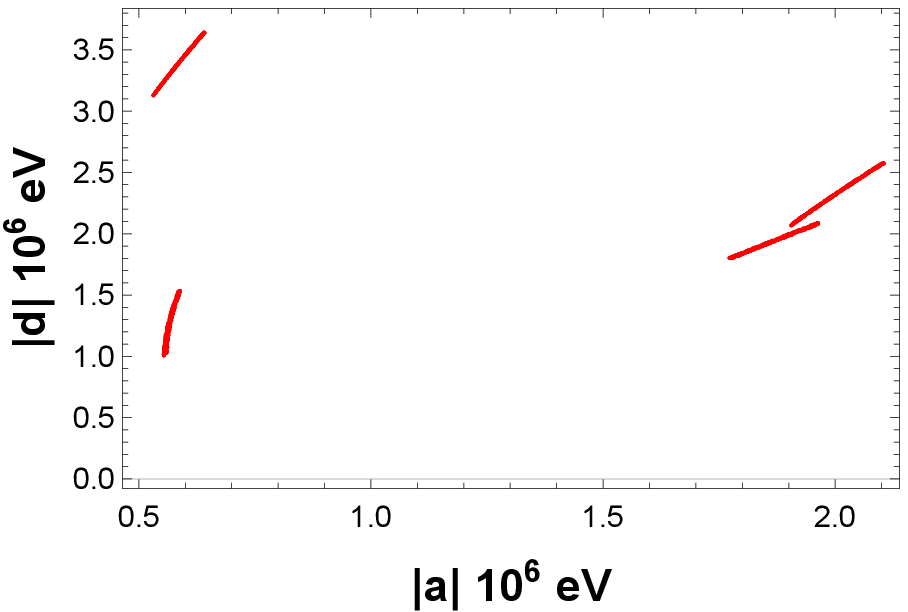}}
\quad
\subfigure[]{
\includegraphics[width=.40\textwidth]{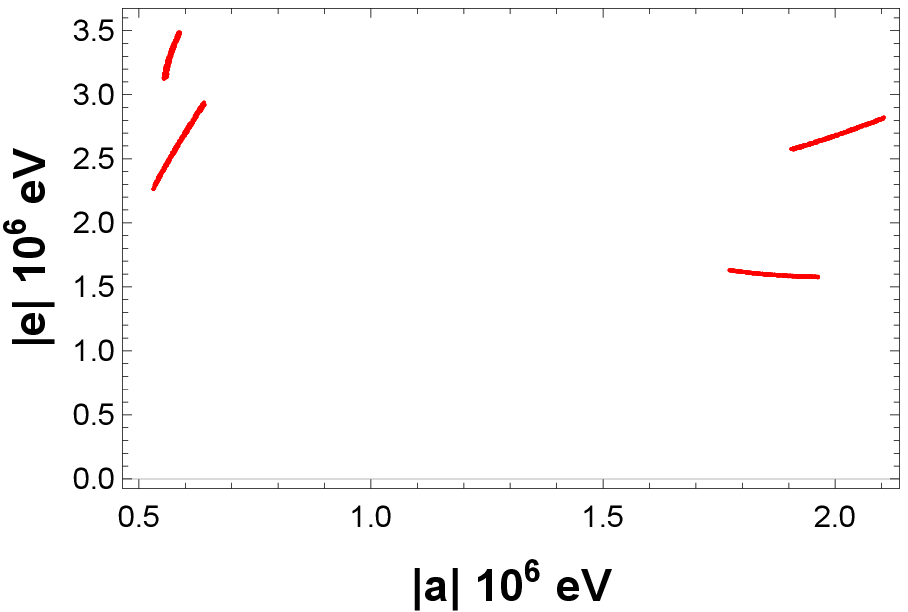}}
\caption{Correlation plots between the model parameters for normal hierarchy (NH).}
\label{f0} 
\end{figure}

\begin{figure}[!h]
\centering
\subfigure[]{
\includegraphics[width=.40\textwidth]{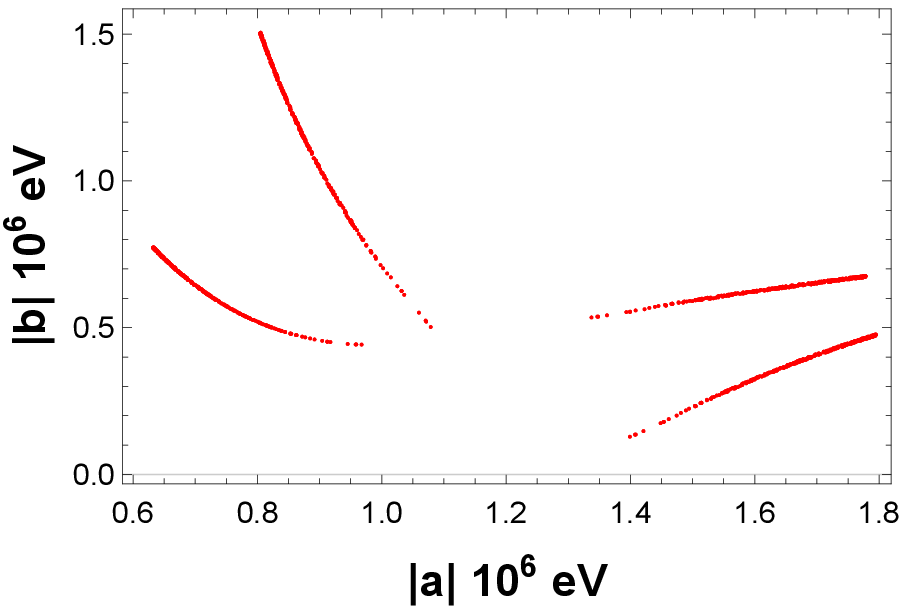}}
\quad
\subfigure[]{
\includegraphics[width=.40\textwidth]{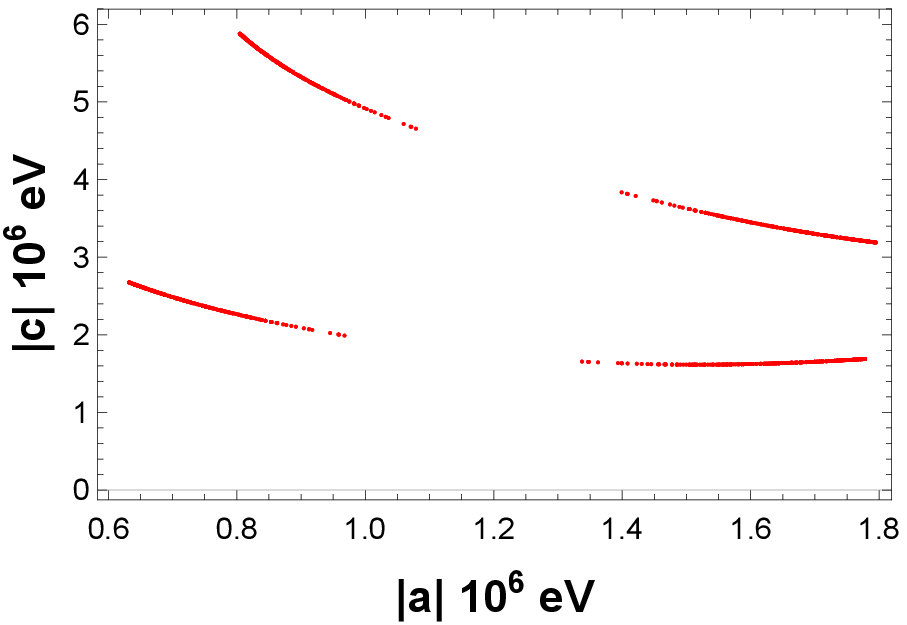}}
\quad
\subfigure[]{
\includegraphics[width=.40\textwidth]{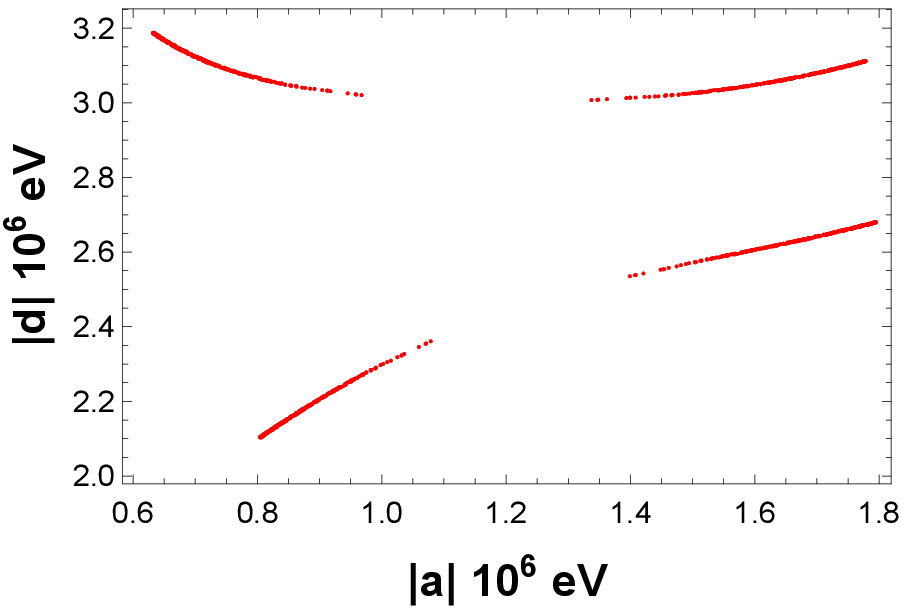}}
\quad
\subfigure[]{
\includegraphics[width=.40\textwidth]{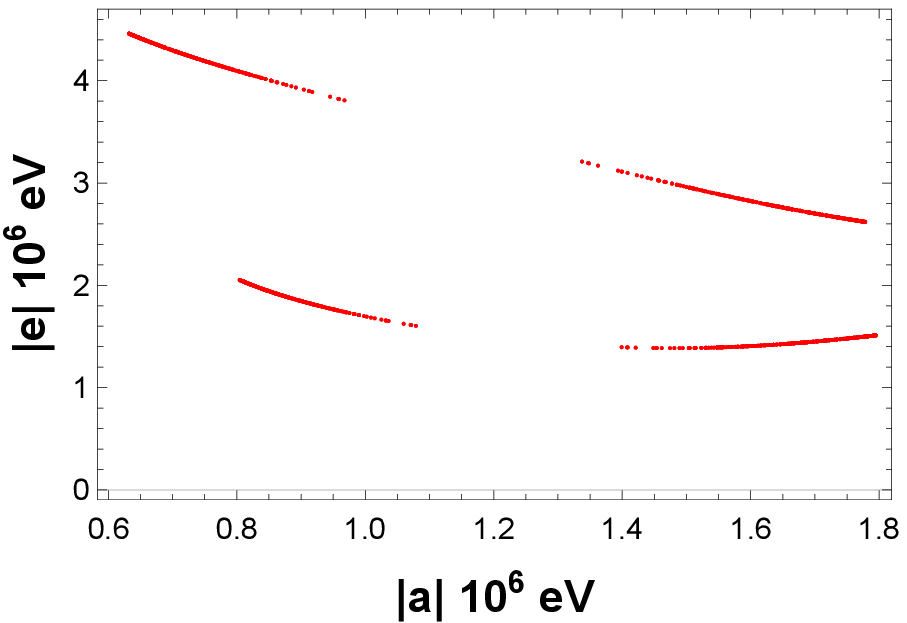}}
\caption{Correlation plots between the model parameters for inverted hierarchy (IH).}
\label{fi}
\end{figure}

\begin{figure}[!h]
\centering
\subfigure[]{
\includegraphics[width=.45\textwidth]{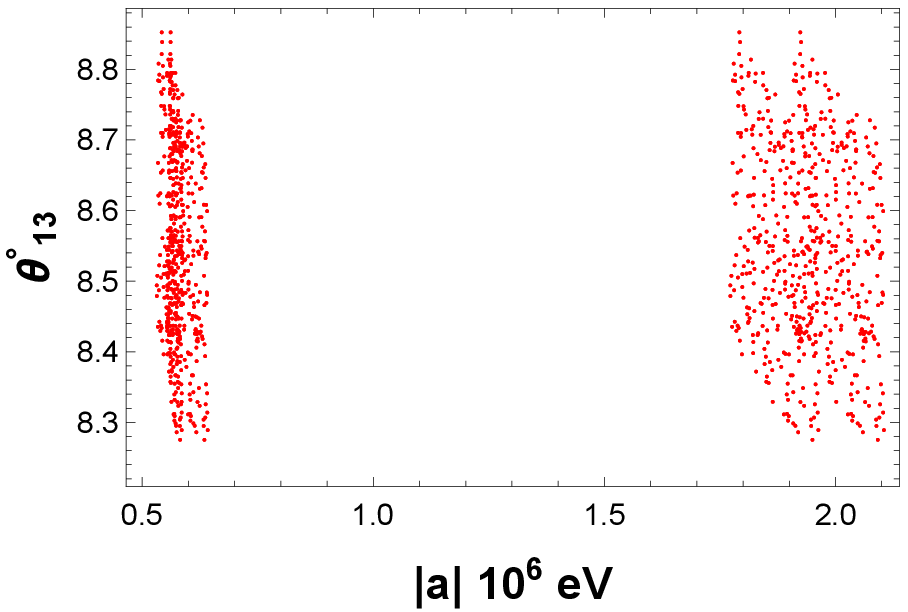}}
\quad
\subfigure[]{
\includegraphics[width=.45\textwidth]{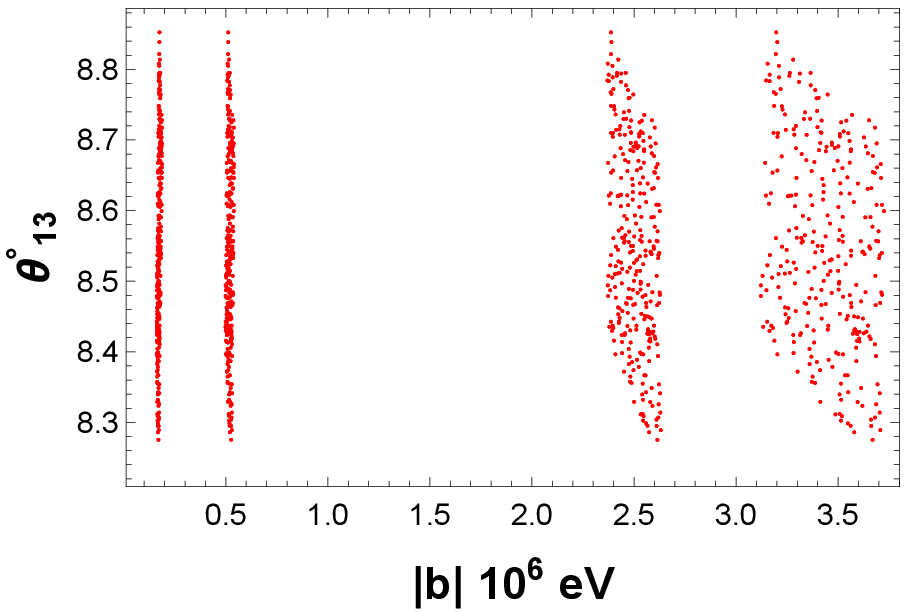}}
\quad
\subfigure[]{
\includegraphics[width=.45\textwidth]{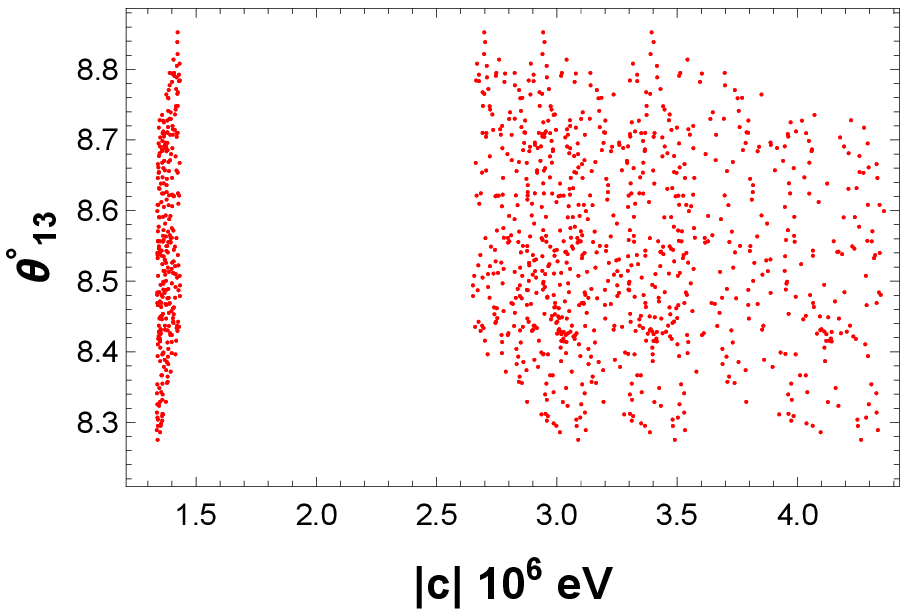}}
\quad
\subfigure[]{
\includegraphics[width=.45\textwidth]{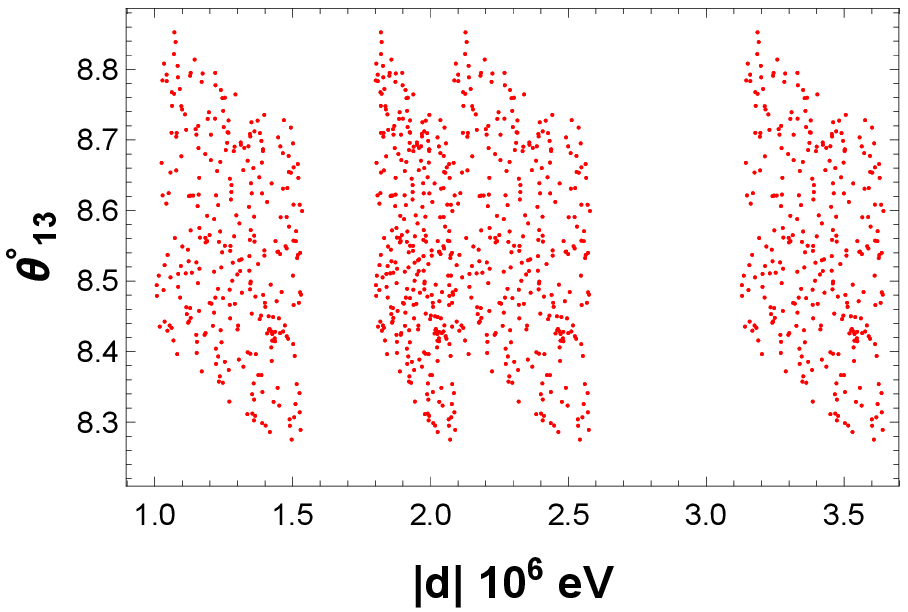}}
\quad
\subfigure[]{
\includegraphics[width=.45\textwidth]{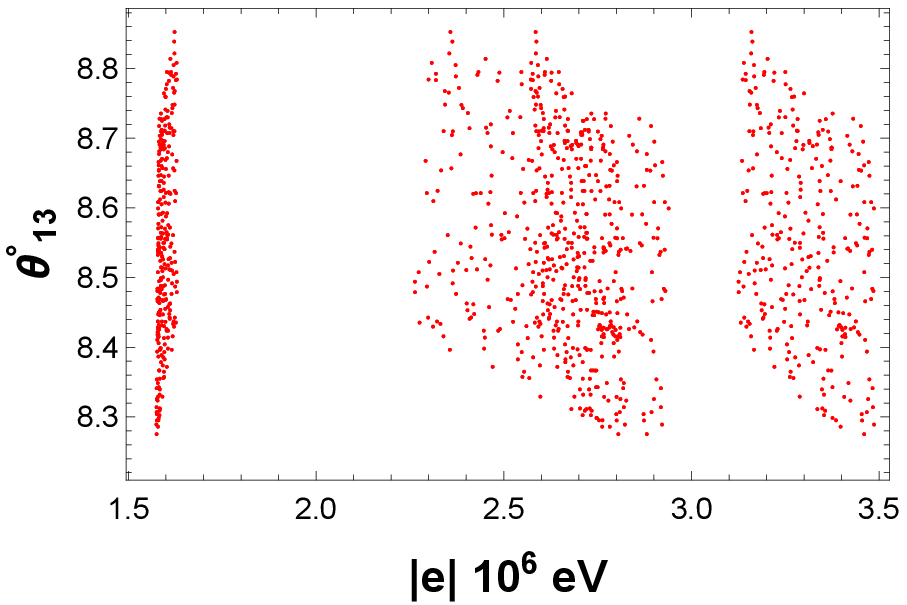}}
\caption{Variation of $\theta_{13}$ with  the model parameters for NH.}
\label{f13n}
\end{figure}
\begin{figure}[!h]
\centering

\subfigure[]{
\includegraphics[width=.45\textwidth]{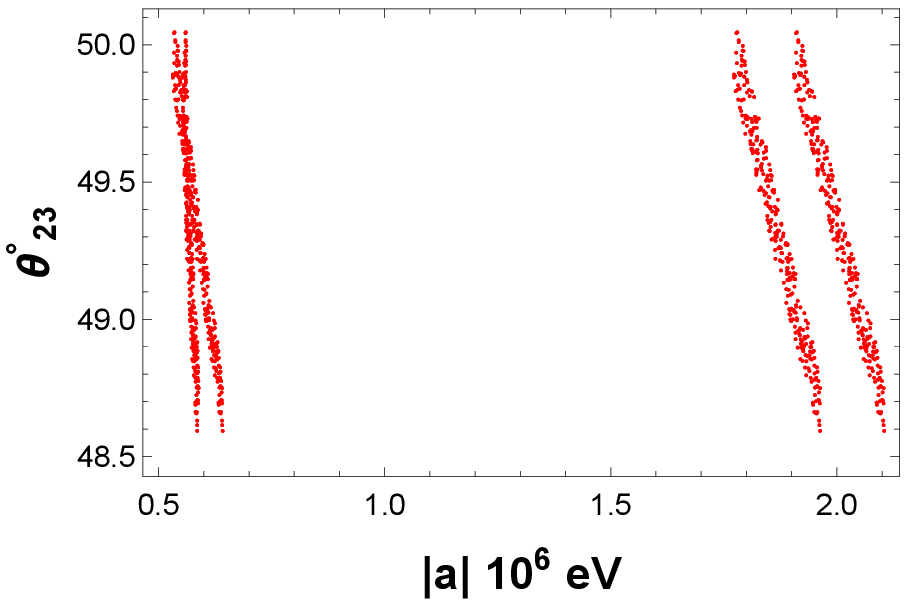}}
\quad
\subfigure[]{
\includegraphics[width=.45\textwidth]{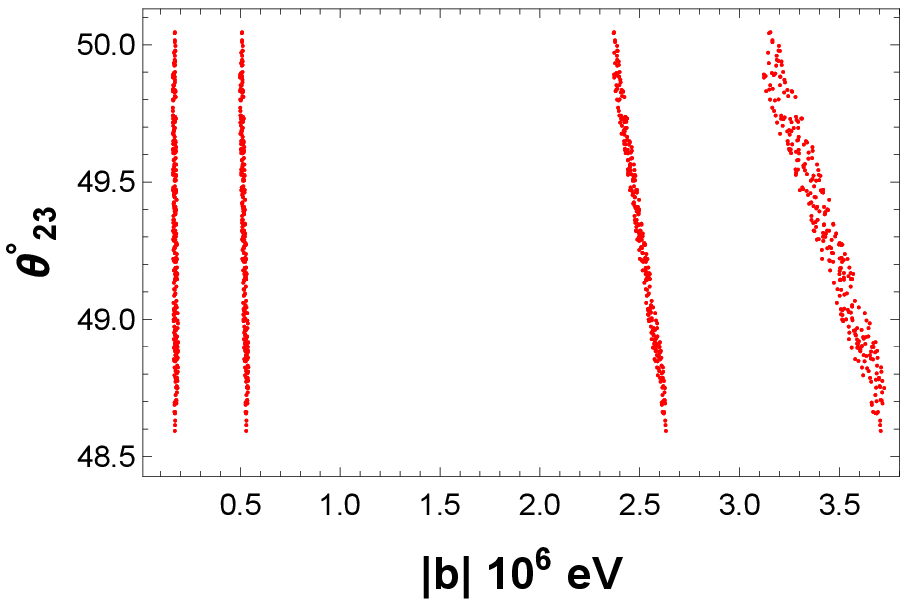}}
\quad
\subfigure[]{
\includegraphics[width=.45\textwidth]{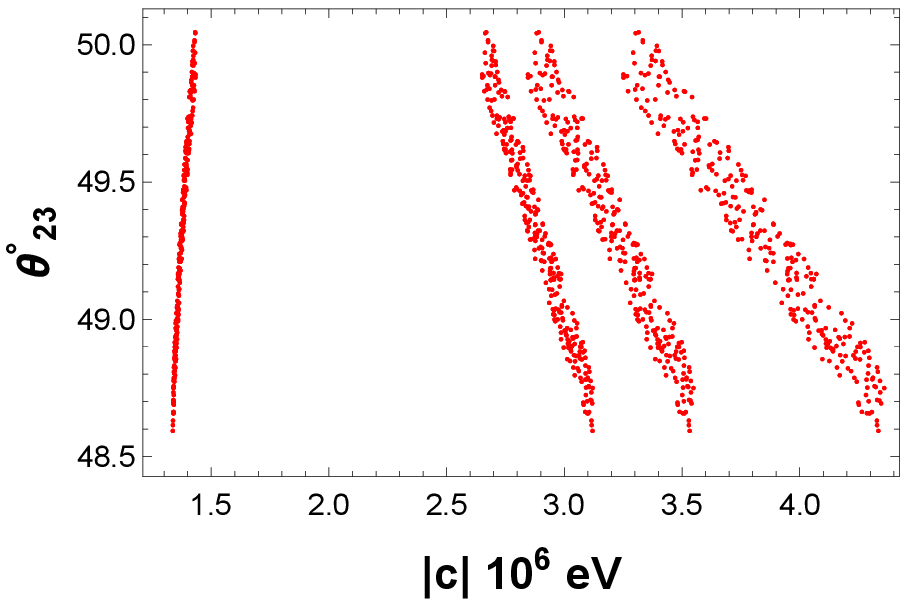}}
\quad
\subfigure[]{
\includegraphics[width=.45\textwidth]{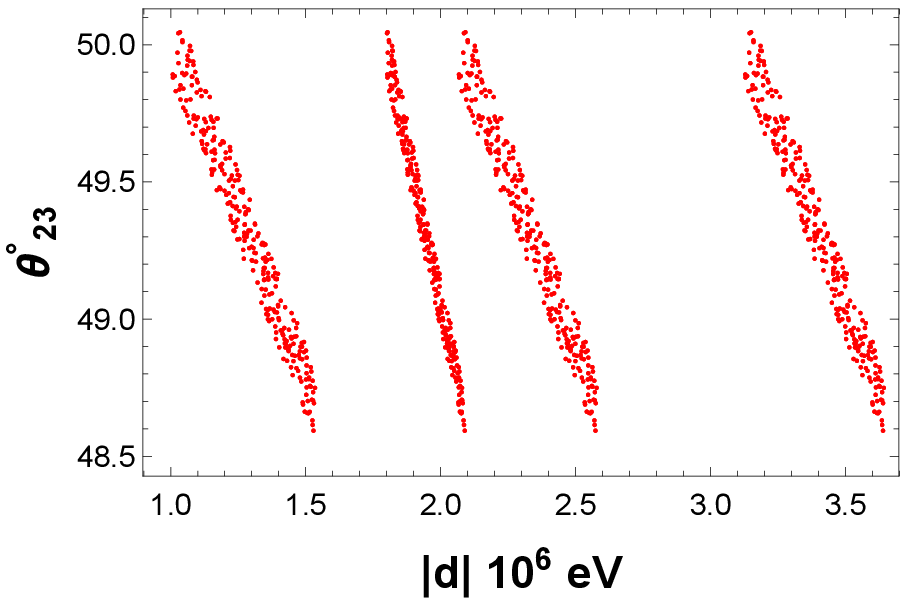}}
\quad
\subfigure[]{
\includegraphics[width=.45\textwidth]{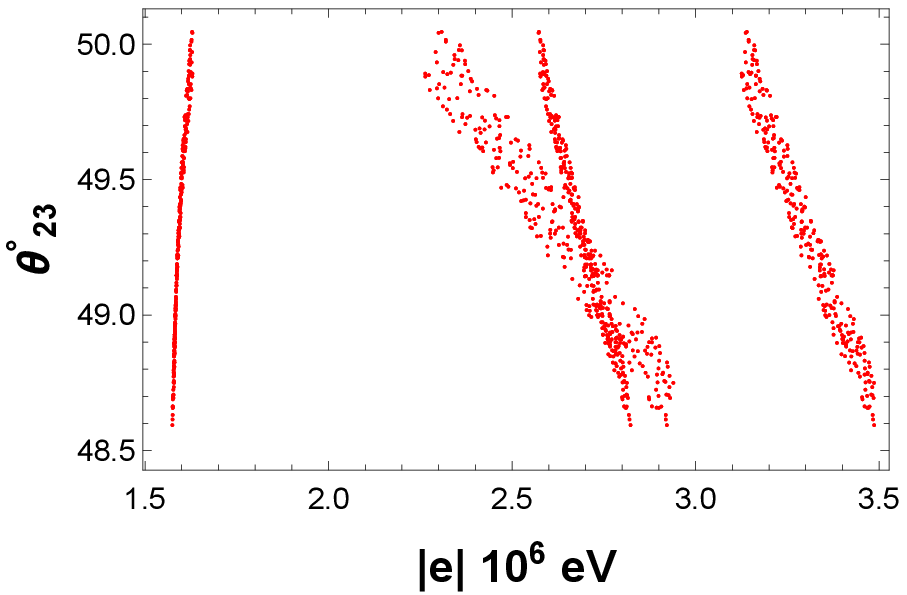}}
\caption{Variation of $\theta_{23}$ with  the model parameters for NH.}
\label{f23n}
\end{figure}
\begin{figure}[!h]
\centering
\subfigure[]{
\includegraphics[width=.45\textwidth]{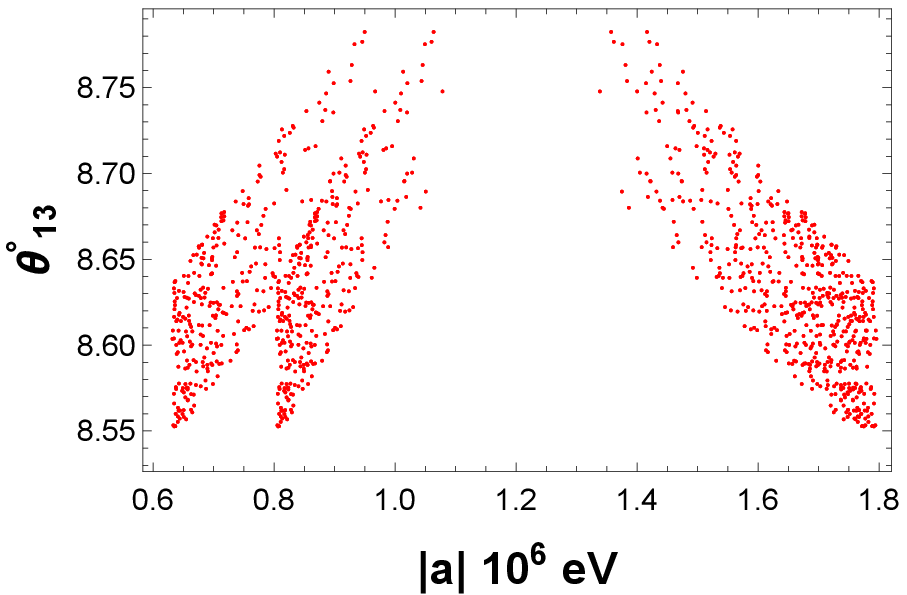}}
\quad
\subfigure[]{
\includegraphics[width=.45\textwidth]{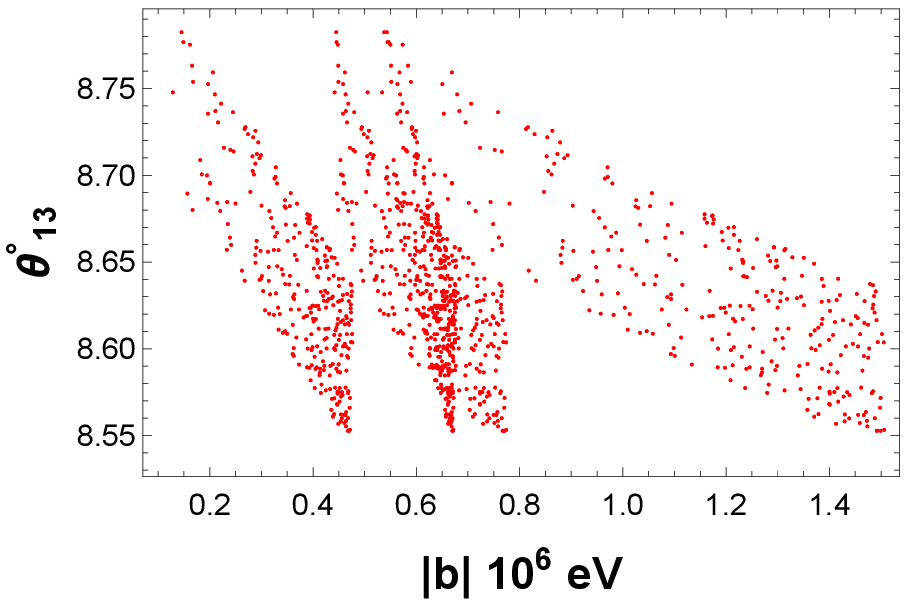}}
\quad
\subfigure[]{
\includegraphics[width=.45\textwidth]{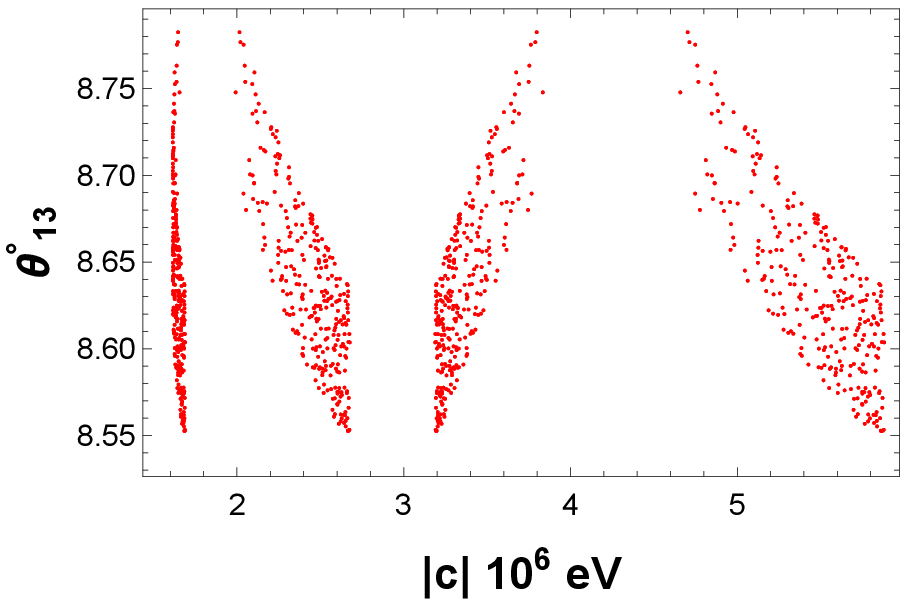}}
\quad
\subfigure[]{
\includegraphics[width=.45\textwidth]{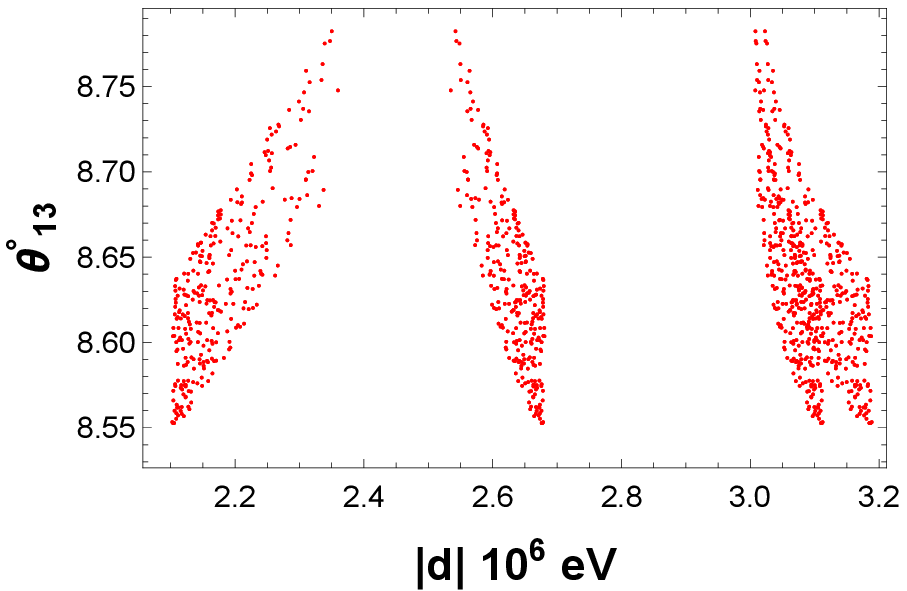}}
\quad
\subfigure[]{
\includegraphics[width=.45\textwidth]{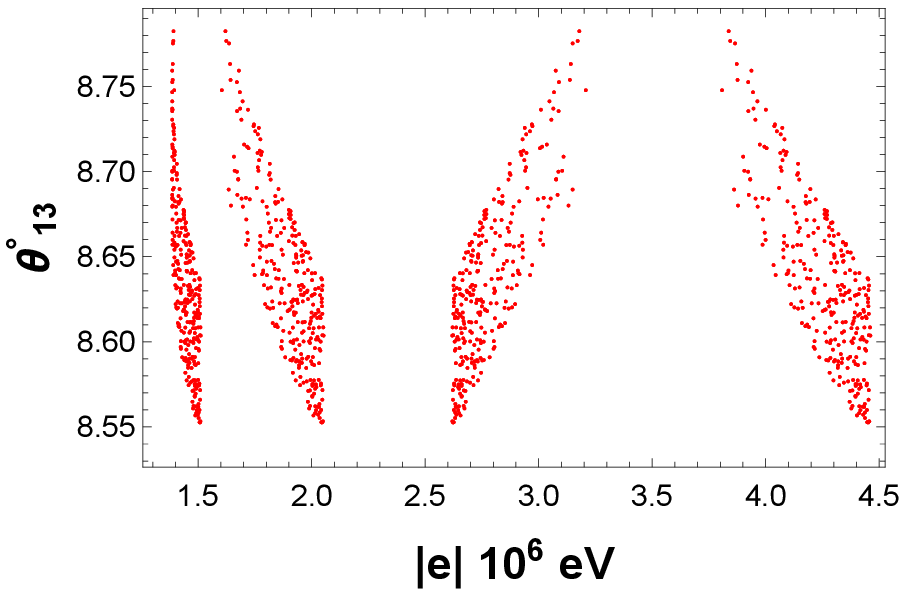}}
\caption{Variation of $\theta_{13}$ with  the model parameters for IH.}
\label{f13i}
\end{figure}
\begin{figure}[!h]
\centering

\subfigure[]{
\includegraphics[width=.45\textwidth]{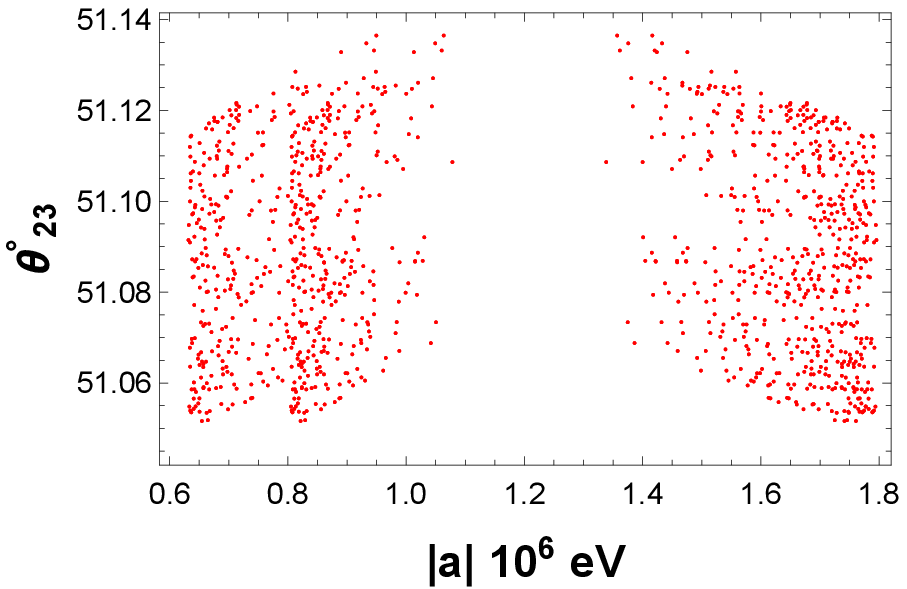}}
\quad
\subfigure[]{
\includegraphics[width=.45\textwidth]{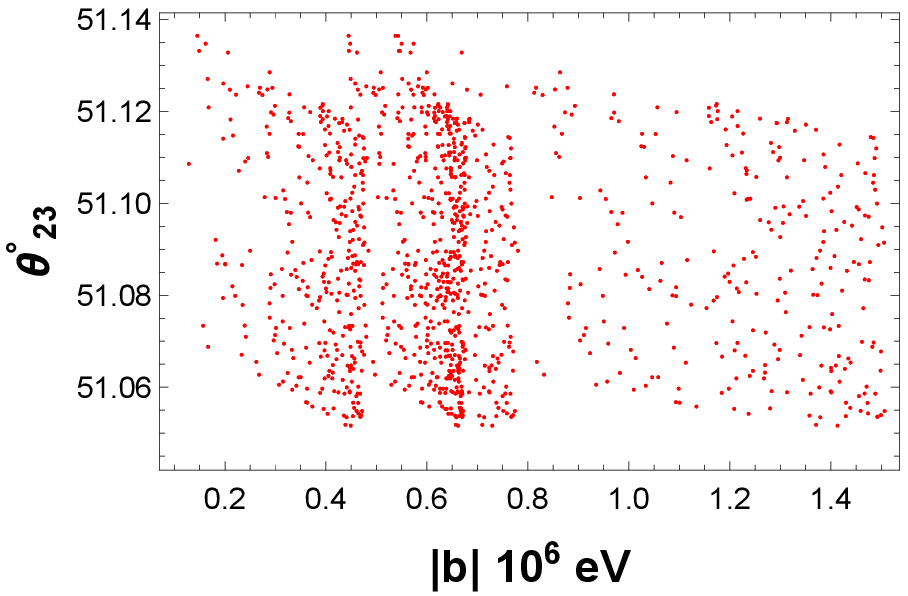}}
\quad
\subfigure[]{
\includegraphics[width=.45\textwidth]{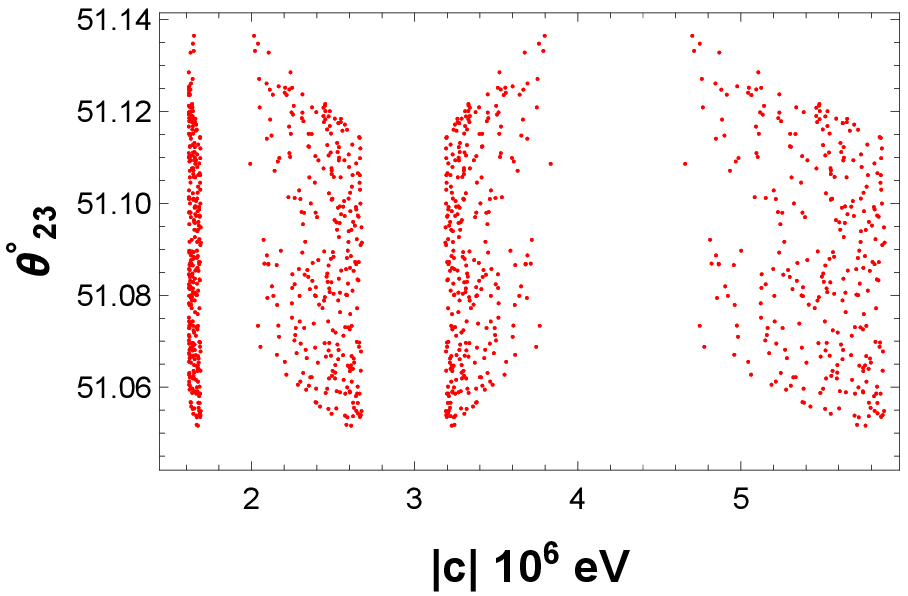}}
\quad
\subfigure[]{
\includegraphics[width=.45\textwidth]{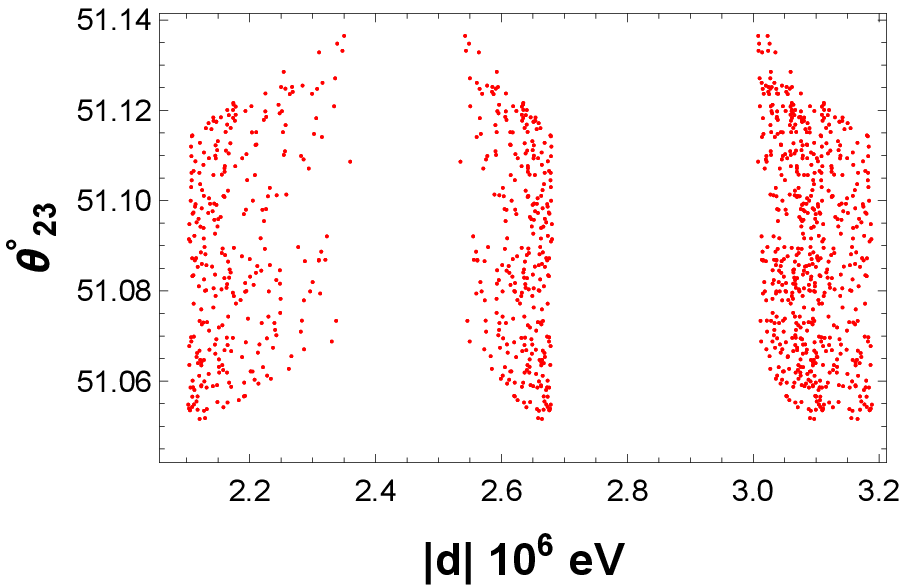}}
\quad
\subfigure[]{
\includegraphics[width=.45\textwidth]{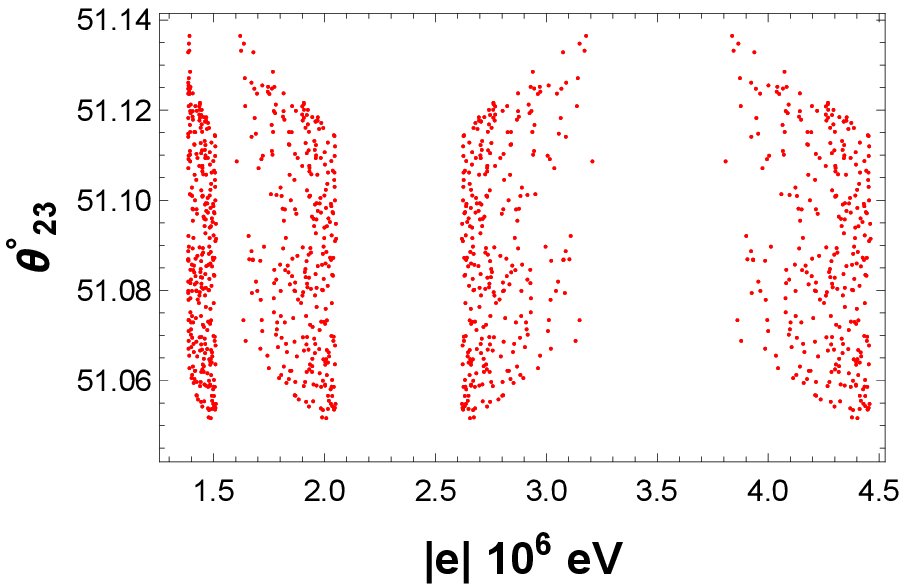}}
\caption{Variation of $\theta_{23}$ with  the model parameters for IH.}
\label{f23i}
\end{figure}
\begin{figure}[!h] 
\centering
\subfigure[]{
\includegraphics[width=.45\textwidth]{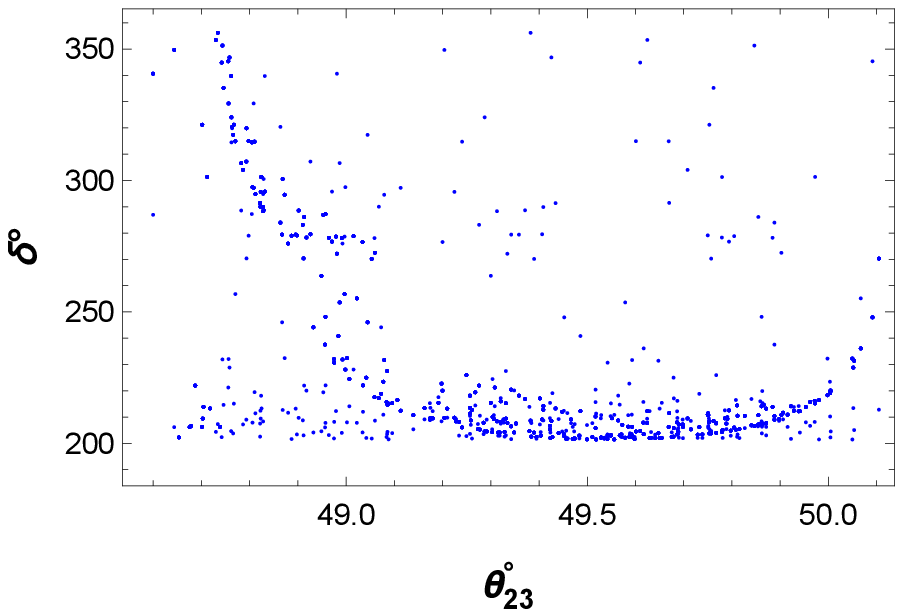}}
\quad
\subfigure[]{
\includegraphics[width=.45\textwidth]{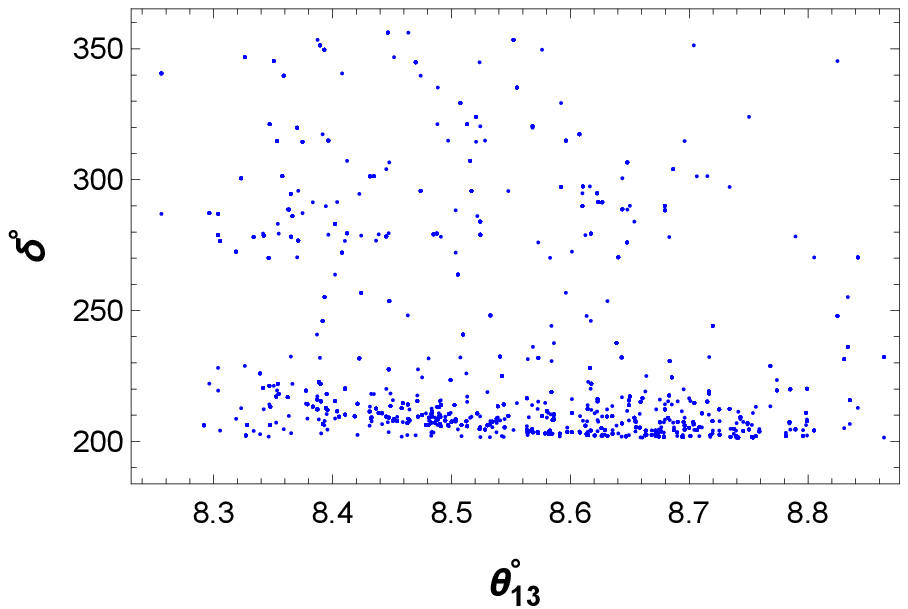}}
\quad
\subfigure[]{
\includegraphics[width=.45\textwidth]{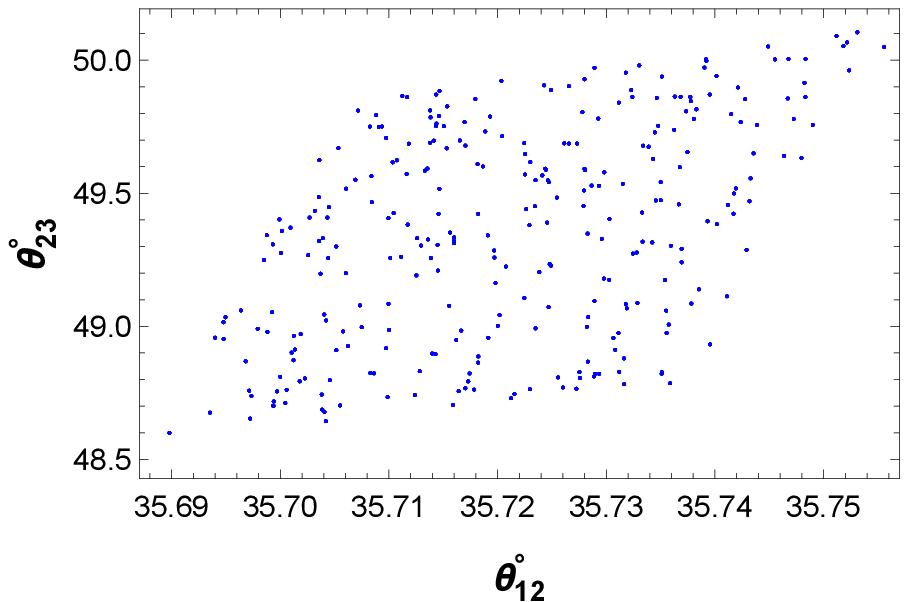}}
\subfigure[]{
\includegraphics[width=.45\textwidth]{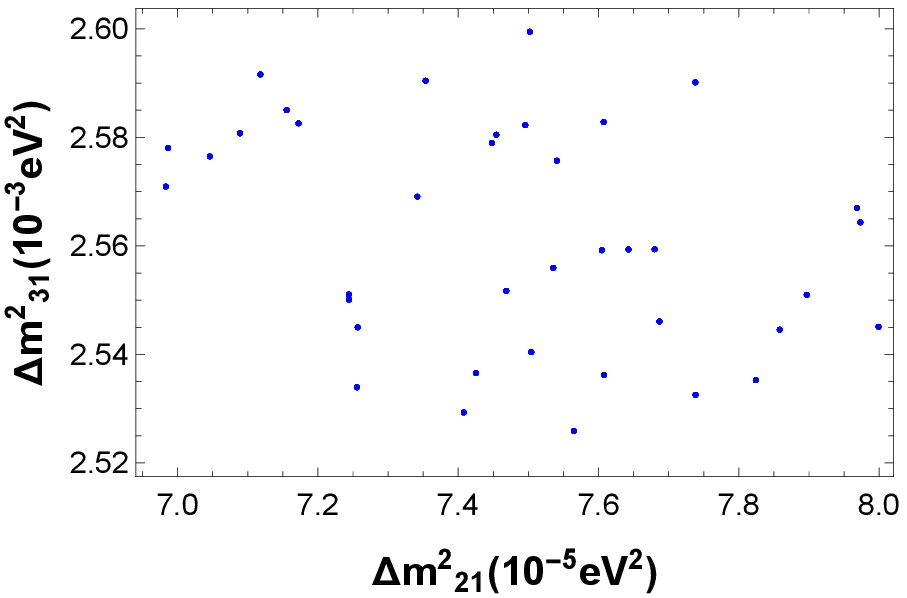}}
\caption{Correlation plots for normal hierarchy (NH). }
\label{f1}
\end{figure}

\begin{figure}[!h]
\centering
\subfigure[]{
\includegraphics[width=.45\textwidth]{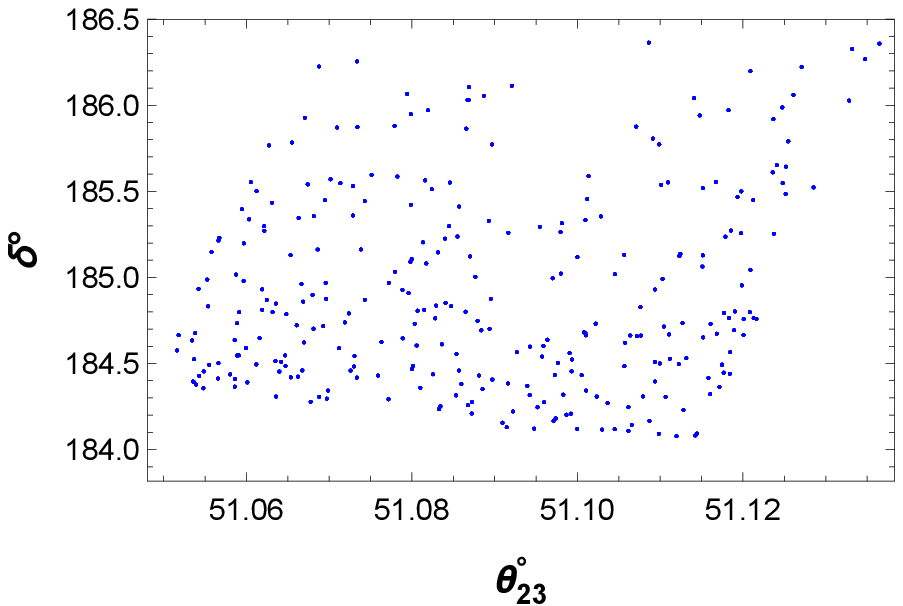}}
\quad
\subfigure[]{
\includegraphics[width=.45\textwidth]{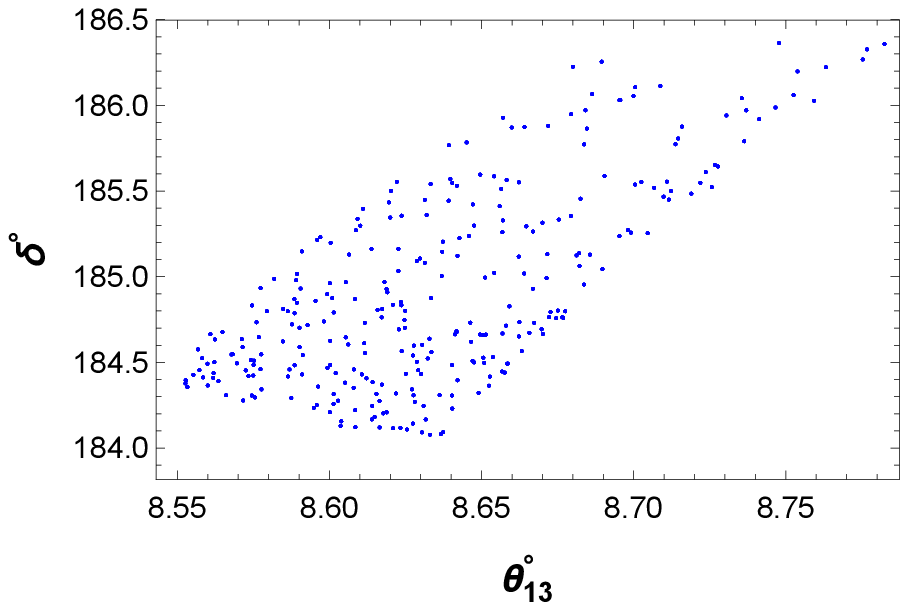}}
\quad
\subfigure[]{
\includegraphics[width=.45\textwidth]{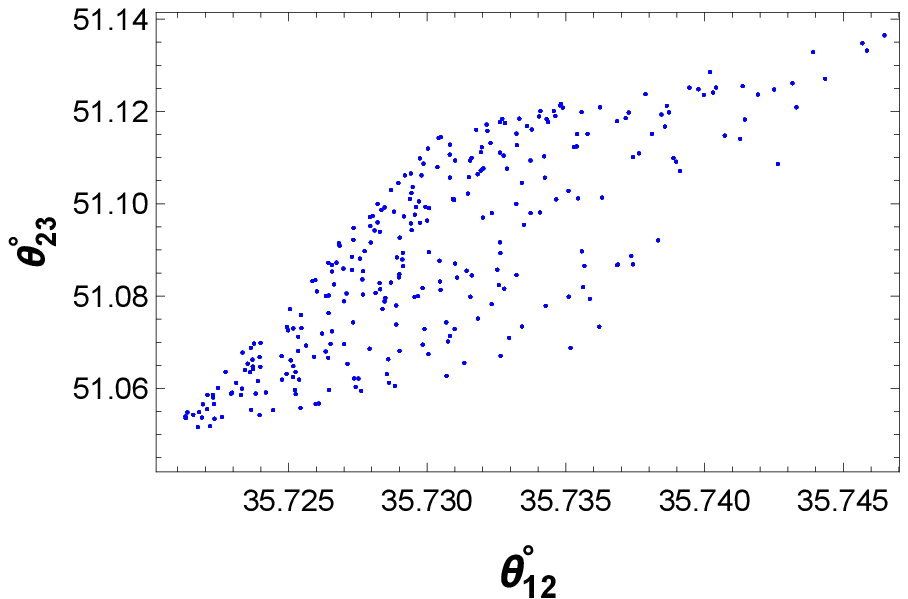}}
\quad
\subfigure[]{
\includegraphics[width=.45\textwidth]{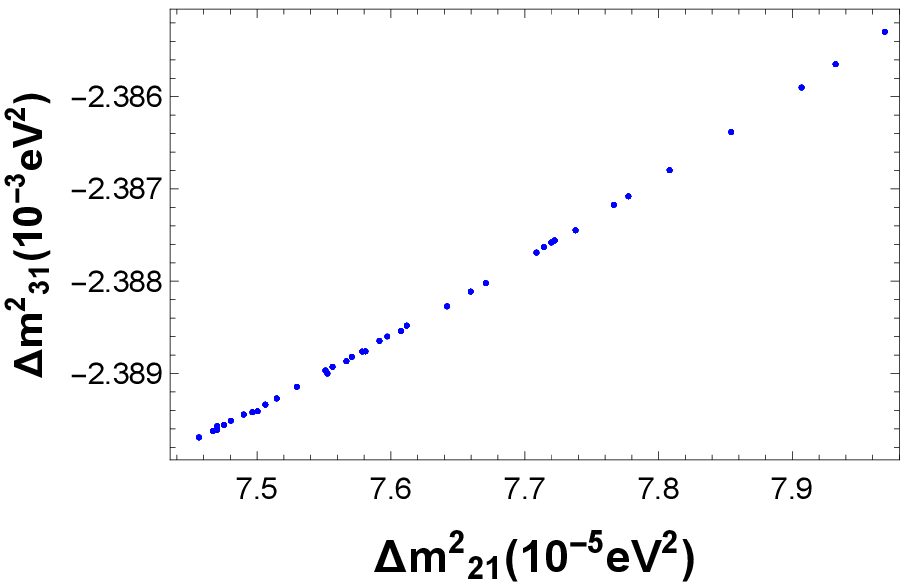}}
\caption{Correlation plots for inverted mass hierarchy (IH).}
\label{f2}
\end{figure}

\begin{figure}[!h]
\centering
\subfigure[]{
\includegraphics[width=.473\textwidth]{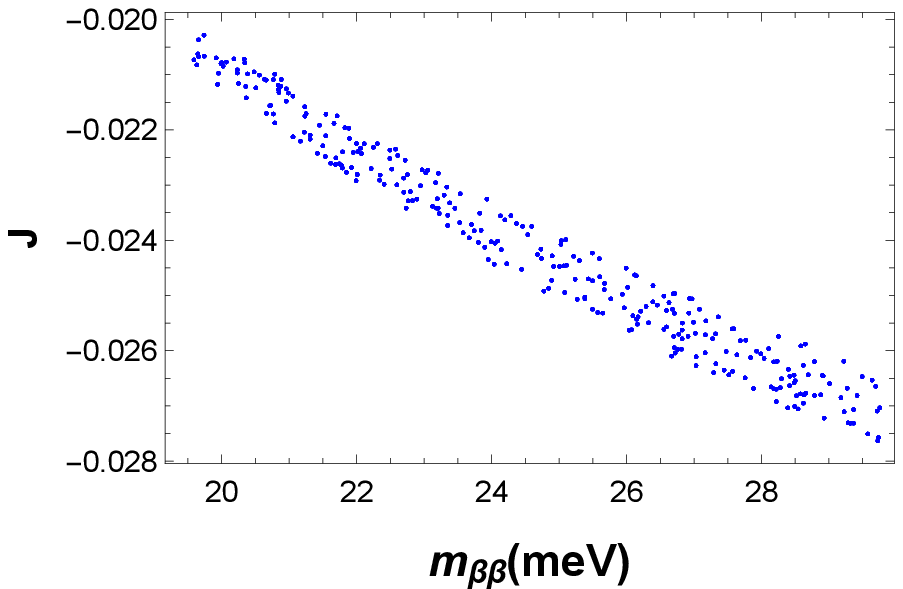}}
\quad
\subfigure[]{
\includegraphics[width=.45\textwidth]{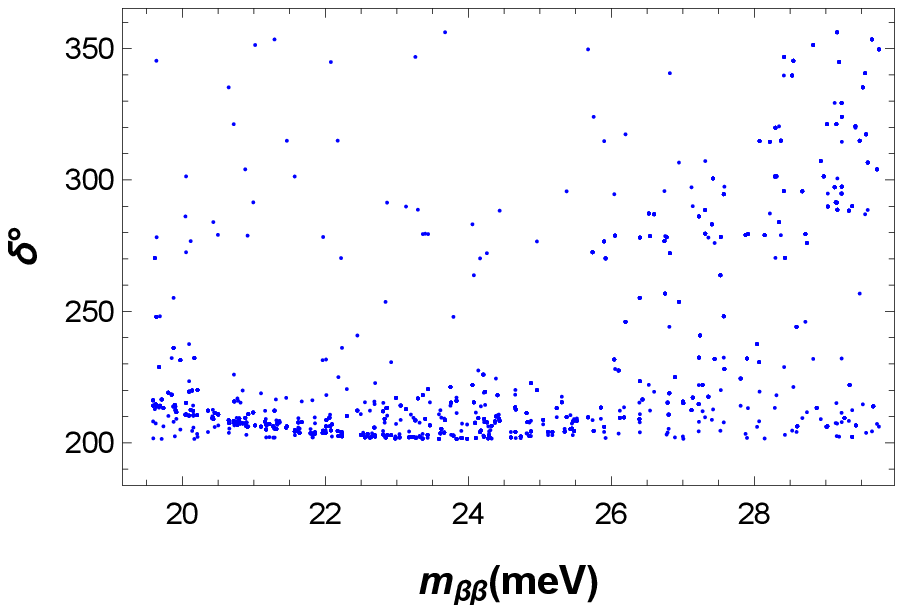}}
\quad
\caption{Model predictions for Jarlskog invariant versus effective Majorana mass and Dirac CP violating phase versus effective Majorana mass for NH. }
\label{f3}
\end{figure}

\begin{figure}[!h]
\centering
\subfigure[]{
\includegraphics[width=.475\textwidth]{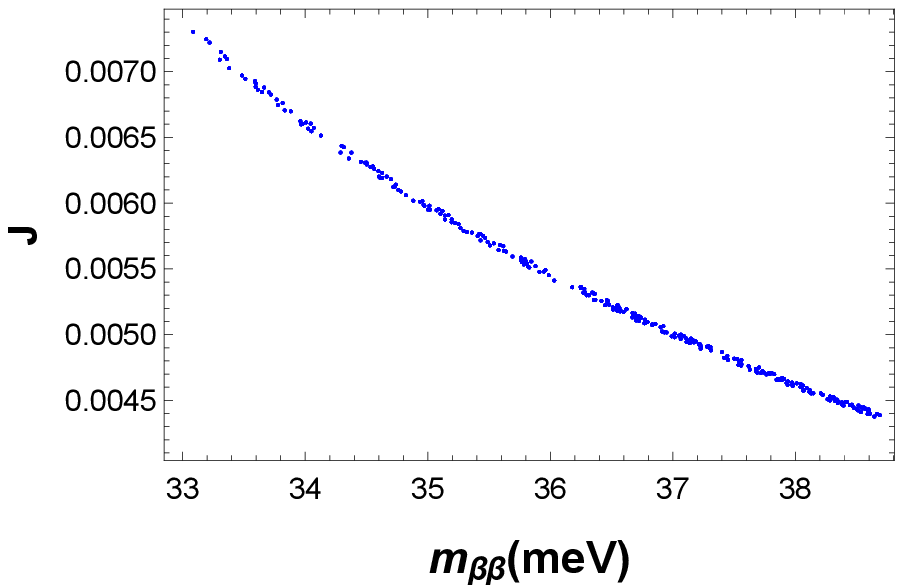}}
\quad
\subfigure[]{
\includegraphics[width=.45\textwidth]{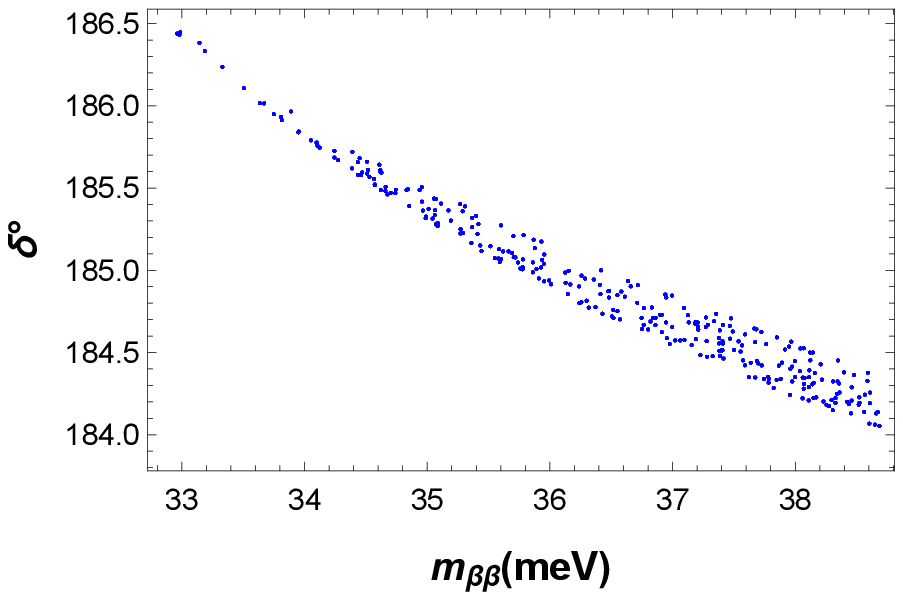}}
\quad
\caption{Model predictions for  Jarlskog invariant versus effective Majorana mass  and Dirac CP violating phase versus effective Majorana mass for IH. }
\label{f4}
\end{figure}
 As the light neutrino mass matrix $m_{\nu}$  obtained in Eq.~(\ref{e4}) is   in the basis where charged lepton mass matrix is diagonal, the Pontecorvo-Maki-Nakagawa-Sakata leptonic mixing matrix ($U_{PMNS}$) which is necessary for the  diagonalization of $m_{\nu}$ becomes a unitary matrix $U$. Therefore, the light neutrino mass matrix $m_{\nu}$ is diagonalized as:
\begin{equation}\label{e5}
m_{\nu}= U^{*}m_{\text{diag}}U^{\dag}
\end{equation}
 where $m_{\text{diag}}=\text{diag}(m_{1},m_{2},m_{3})$ is the  light neutrino mass matrix in the diagonal form.  
 
 The three neutrino mass eigenvalues can be written as,
 \begin{align}
 \label{em}
m_{\text{diag}}=\begin{cases}
\text{diag}(m_{1},\sqrt{m_{1}^{2}+\Delta m_{21}^{2}},\sqrt{m_{1}^{2}+\Delta m_{31}^{2}})~~ &\text{in normal hierarchy (NH)} \\
\text{diag}( \sqrt{m_{3}^{2}+\Delta m_{31}^{2}}, \sqrt{m_{3}^{2}+\Delta m_{31}^{2}+\Delta m_{21}^{2}},m_{3})~~ &\text{in inverted hierarchy (IH)}
\end{cases}
 \end{align}
 
 The upper bound on the sum of neutrino masses ($\sum m_{i}=m_{1}+m_{2}+m_{3})$ obtained by the Planck is 0.12 eV \cite{Planck:2018nkj}. 
 
  The PMNS matrix U can be parametrized in terms of neutrino mixing angles and Dirac CP phase $\delta$. Following the PDG convention \cite{ParticleDataGroup:2018ovx}, U takes the form 
 \begin{equation}
 U_{\text{PMNS}} = P_{\phi} 
	\begin{pmatrix}
	c_{12} c_{13} & s_{12} c_{13} & s_{13}e^{-i\delta} \\
	-s_{12} c_{23} - c_{12} s_{23} s_{13}e^{i\delta} & c_{12}c_{23} - s_{12} s_{23} s_{13}e^{i\delta} & s_{23} c_{13} \\
	s_{12} s_{23} - c_{12} c_{23} s_{13}e^{i\delta} & -c_{12} s_{23} - s_{12} c_{23} s_{13}e^{i\delta} & c_{23} c_{13}
	\end{pmatrix}
	\begin{matrix}
	  P
	\end{matrix},
 \end{equation}
   where $\theta_{ij}$ (for $\text{ij}= 12,$ 13, 23)  are  the  mixing  angles  (with $c_{ij}=\cos{\theta}_{ij}$ and $s_{ij}=\sin{\theta}_{ij}$. $P=\text{diag}(e^{i\alpha},e^{i\beta},1)$ contains  two  Majorana CP  phases $\alpha$ and $\beta$,  while $P_{\phi} = \text{diag}(e^{i\phi_{1}}, e^{i\phi_{2}}, e^{i\phi_{3}})$ consists of three unphysical phases $\phi_{1,2,3}$ that can be removed via the charged-lepton field rephasing \cite{Xing:2015fdg}.  The neutrino  mixing angles $\theta_{12}$, $\theta_{23}$ and $\theta_{13}$ in terms of the elements of U are  given below: 
   \begin{align}
   s_{12}^{2}=\frac{|U_{e2}|^{2}}{1-|U_{e3}|^{2}},~~ s_{23}^{2}=\frac{|U_{\mu 3}|^{2}}{1-|U_{e3}|^{2}},~~ s_{13}^{2}=|U_{e3}|^{2}
   \label{eu}
   \end{align}
    
The Jarlskog invariant is given by the phase redefinition invariant quantity,
 \begin{equation}
 J=\text{Im}\{U_{e1}U_{\mu 2}U_{e2}^{*}U_{\mu 1}^{*}\} =s_{12}c_{12}s_{23}c_{23}c_{13}^{2}s_{13}\sin \delta.
 \end{equation}
 
  In order to show that the model is  consistent with the present neutrino oscillation data,  we vary the free parameters of our model $a,~b,~c,~d~\text{and}~e$ to fix the neutrino oscillation observables  $\theta_{13}$,~$\theta_{23}$,~$\theta_{12}$,~$\Delta m_{21}^{2}$,~$\Delta m_{31}^{2}$ and $\delta$ to their experimental values. The experimental values of neutrino oscillation  parameters used in our analysis are given in Table \ref{t3}. The allowed regions of the model parameters that can satisfy current oscillation data for both NH and IH are shown in   Fig.~\ref{f0} and Fig.~\ref{fi}, respectively in the form of correlation plots.

The values of $\theta_{13}$  obtained in the allowed regions of our model parameters for both NH and IH are shown in Fig.~\ref{f13n} and  Fig.~\ref{f13i}, respectively while Fig.~\ref{f23n} and Fig.~\ref{f23i} give the variation of $\theta_{23}$ in NH and IH, respectively. The model predictions of the neutrino oscillation parameters in 3$\sigma$ range are shown in Fig.~\ref{f1} and Fig.~\ref{f2}. One of the important predictions of model is that  solar neutrino mixing angle $\theta_{12}$ lies around $35.73^{\circ}$ in both NH and IH cases. In NH, the Dirac CP violating phase $\delta$ is obtained in the range $201.47^{\circ} \leq\delta \leq 356.20^{\circ}$ as shown in Fig.~\ref{f1}, with  the average model value of $230.89^{\circ}$. The reactor angle $\theta_{13}$ and atmospheric angle  $\theta_{23}$ predicted by the model are outside 1$\sigma $ range but well within 3$\sigma$ range. In the case of $\theta_{23}$, our present model prefers the higher octant ($>45^{\circ}$), with the average model value of $ 49.32^{\circ}$.  The average model value of $\Delta m^{2}_{21}$ and  $\Delta m^{2}_{31}$ are $7.46\times 10^{-5}\text{eV}^{2}$ and $2.56\times 10^{-3}\text{eV}^{2}$, respectively and show good agreement with the experimental data.

 The predictions in inverted hierarchy are shown in Fig.~\ref{f2}. In this case, the predicted values for $\theta_{13}$, $\Delta m^{2}_{21}$ and $\Delta m^{2}_{31}$ are all in the 3$\sigma$ range with the average model values of $8.63^{\circ}$, $7.61\times 10^{-5}\text{eV}^{2}$ and $-2.38\times 10^{-3}\text{eV}^{2}$, respectively. The angle $\theta_{23}$ are  predicted in the higher octant ($>45^{\circ}$) with the average model value of $51.09^{\circ}$. The Dirac CP violating phase $\delta$ is predicted in the range  $184.05^{\circ} \leq \delta \leq 186.45^{\circ}$ and shows a deviation from the global fit values. The current analysis is consistent with the latest cosmological bound  $\Sigma m_{i}\leq$ 0.12 eV. Thus, the presented model  produces the required deviation from TBM necessary to accommodate the current neutrino oscillation data with distinctive predictions of Dirac CP violating phase $\delta$ in IH. 

 The additional predictions for the effective Majorana mass $|m_{\beta \beta}|$ vs Jarlskog invariant J in 3$\sigma$ range for both NH and IH are shown in Fig.~\ref{f3}(a) and Fig.~\ref{f4}(a), respectively. The correlation between $\delta$ and $|m_{\beta \beta}|$ are depicted in Fig.~\ref{f3}(b) for NH and Fig.~\ref{f4}(b) for IH.  
 
 And, the effective Majorana mass $|m_{\beta \beta}|$ is given by
 \begin{equation}
 |m_{\beta \beta}|=|\sum\limits_{i} U_{ei}^{2}m_{i}|.
\end{equation}
 The upper limits of the effective Majorana mass are obtained by:  Gerda \cite{GERDA:2019ivs} as $|m_{\beta \beta}|<(104-228)$ meV corresponds to                                                                                     $^{76}Ge\ (T^{0\nu \beta \beta}_{1/2}>9\times 10^{25}$ yr),  CUORE \cite{CUORE:2019yfd} as $|m_{\beta \beta}|<(75-350)$ meV corresponds to                                                                                    $^{130}Te\ (T^{0\nu\beta \beta}_{1/2}>3.2\times 10^{25}$ yr) and  KamLAND-Zen \cite{KamLAND-Zen:2016pfg} as $|m_{\beta \beta}|<(61-165)$ meV corresponds to $^{136}Xe\ (T^{0\nu \beta \beta}_{1/2}>1.07\times 10^{25}$ yr).  Our predicted $3\sigma$ range values of $| m_{\beta \beta}|$ are $(19.59-29.75)$ meV in NH and $(32.96-38.69)$ meV in IH . And the predicted range of $|m_{\beta \beta}|$ for both cases can be tested  in  future as there are planned ton-scale and next generation $0\nu \beta \beta$ experiments using $^{136}Xe$ \cite{KamLAND-Zen:2012vpv, EX}  and $^{76}Ge$ \cite{Abt:2004yk, Majorana:2011rit} that can reach a sensitivity of $|m_{\beta \beta}| \sim (12-30)$ meV, corresponding to $T^{0\nu \beta \beta}_{1/2} \geq 10^{27}$ yr \cite{CarcamoHernandez:2017kra, Lei:2020nik}.      
 
\section{Summary and Conclusion}

In summary, we have presented a neutrino mass model using $A_{4}$ discrete symmetry.  The model uses five extra SM singlets to produce the required deviation from TBM necessary to accommodate current neutrino oscillation data. The model has some characteristic  predictions for neutrino oscillation parameters. The solar neutrino angle $\theta_{12}$ is centred  around $35.73^{\circ}$ for both the mass ordering. The predicted range of the atmospheric neutrino mixing angle $\theta_{23}$ for NH is in good agreement with the experimental data, with the average model value of $49.32^{\circ}$. The average model values of angle $\theta_{13}$ and $\theta_{23}$ are  $8.63^{\circ}$ and  $51.09^{\circ}$ for inverted hierarchy. The predicted range for $\delta$ for IH shows deviation from the of global fit, with the average  model value of $184.88^{\circ}$. The presented model slightly prefers the NH data. The predicted range of  $|m_{\beta \beta}|$ for both  NH and IH can be tested in the near future.

\section*{Acknowledgements}

One of us (VP)  wishes to thank  Department of Science and Technology (DST), Government of India for providing INSPIRE Fellowship. We are thankful to Prof. M.K. Das, Department of Physics, Tezpur University, Assam, for fruitful discussions.

 \appendix  
\section{$A_{4}$ Group}

$A_{4}$ is the even permutation group of 4 objects with $\frac{4!}{2}$ elements. It has four irreducible representations, namely 1, $1'$, $1''$ and 3. All the elements of the group can be generated by two elements S and T. The generators S and T satisfy the relation,
\begin{equation}
S^{2}=(ST)^3=T^3=1.
\end{equation}

 The multiplication rules of any two irreducible representations  under $A_{4}$  are given by
$$ 3\otimes 3=1\oplus 1^{'}\oplus 1^{''}\oplus 3_{S}\oplus  3_{A}$$ 
\begin{equation}
\begin{array}{ccc}
1\otimes1=1 \hspace{2cm}& 1'\otimes1'=1''\\
1"\otimes1''=1' \hspace{2cm} &  1'\otimes1''=1\\
3\otimes 1/1'/1''= 3\hspace{2cm}&1/1'/1''\otimes3=3
\end{array}.
\end{equation}
where
\begin{equation}
1=a_{1}b_{1}+a_{2}b_{3}+a_{3}b_{2}
\end{equation}
\begin{equation}
1^{'}=a_{3}b_{3}+a_{1}b_{2}+a_{2}b_{1}
\end{equation}
\begin{equation}
1^{''}=a_{2}b_{2}+a_{1}b_{3}+a_{3}b_{1}
\end{equation}
\begin{equation}
3_{S}=\begin{pmatrix}
2a_{1}b_{1}-a_{2}b_{3}-a_{3}b_{2}\\
2a_{3}b_{3}-a_{1}b_{2}-a_{2}b_{1}\\
2a_{2}b_{2}-a_{1}b_{3}-a_{3}b_{1}
\end{pmatrix}
\end{equation}
\begin{equation}
3_{A}=\begin{pmatrix}
a_{2}b_{3}-a_{3}b_{2}\\
a_{1}b_{2}-a_{2}b_{1}\\
a_{3}b_{1}-a_{1}b_{3}
\end{pmatrix}
\end{equation}

  The detailed studies on $A_{4}$ symmetry can be found in Ref \cite{Ishimori:2010au}.

\bibliographystyle{ieeetr}
\bibliography{vreff}
\end{document}